\documentclass[preprint,epsfig,floats,aps]{revtex4}

\usepackage{epsfig}
\usepackage{graphicx}

\begin{document}
\title{Discrepancy between hadron matter and quark-gluon matter in net charge 
       transfer fluctuation}
\author{Dai-Mei Zhou$^1$, Xiao-Mei Li$^2$, Bao-Guo Dong$^2$, and 
        Ben-Hao Sa$^{3,2,1,4}$ \footnote{E-mail: sabh@iris.ciae.ac.cn}}
\affiliation{
$^1$  Institute of Particle Physics, Huazhong Normal University,
      Wuhan, 430079 China \\
$^2$  China Institute of Atomic Energy, P. O. Box 275 (18),
      Beijing, 102413 China \\
$^3$  CCAST (World Lab.), P. O. Box 8730 Beijing, 100080 China\\
$^4$  Institute of Theoretical Physics, Academy Sciences, Beijing,
      100080 China}
\begin{abstract}
A parton and hadron cascade model, PACIAE, is employed to investigate the net charge 
transfer fluctuation within $|\eta|$=1 in $Au+Au$ collisions at $\sqrt{s_{NN}}$=200 
GeV. It is turned out that the observable of net charge transfer fluctuation, $\kappa$
, in hadronic final state (HM) is nearly a factor of 3 to 5 larger than that in 
initial partonic state (QGM). However, only twenty percent of the net charge transfer 
fluctuation in the QGM can survive the hadronization. \\ 
  
\noindent{PACS numbers: 25.75.Dw, 24.85.+p, 24.10.Lx}
\end{abstract}
\maketitle

Recently the net charge transfer fluctuation has been proposed in \cite{jeon1,jeon2} 
as a signal of Quark-Gluon-Plasma (QGP) phase transition expected to be existing in 
the relativistic nucleus-nucleus collisions. Following \cite{quig} the observable 
\begin{equation}
\kappa(\eta)=D_u(\eta)/(dN_{ch}/d{\eta})
\end{equation} 
is employed to describe the net charge transfer fluctuation. In the above equation the 
net charge transfer deviation, $D_u$, reads
\begin{equation}
D_u(\eta)=<u(\eta)^2>-<u(\eta)>^2,
\end{equation}
where the net charge transfer, $u$, is defined by
\begin{equation}
u(\eta)=[Q_F({\eta})-Q_B({\eta})]/2,
\end{equation}
where the $Q_F({\eta})$ ($Q_B({\eta})$) is referred to the net charge in forward 
(backward) region of $\eta$ and the $N_{ch}$ stands for the charge multiplicity 
accounted according to the charge of particle. The $\kappa$ is argued to be a measure 
of the local unlike-sign charge correlation length \cite{jeon2} and the charge 
correlation length in QGP phase (in quark-gluon matter, QGM) is expected to be much 
smaller than the one in hadronic matter (HM) because the charge unit is 1/3 and 1 in 
QGM and HM \cite{jeon3,bass1,asak}, respectively.   

In \cite{jeon2} a neutral cluster model was used first and the hadronic transport 
models of HIJING \cite{wang1}, RQMD \cite{sorg1}, and UrQMD \cite{bass2} were  
employed then to study the net charge transfer fluctuation. The results from above 
hadronic transport models could be summarized as follows: 1. The discrepancy among 
them is not obvious from each other. 2. The $\kappa(\eta=1)$ calculated in interval of 
$|\eta|<1$ is equivalent to the net charge fluctuation at $\eta=1$ and is close to the 
STAR datum \cite{star1} of $\sim0.27\pm0.02$ (cited from \cite{jeon2} directly) in 
$Au+Au$ collisions at $\sqrt{s_{NN}}$=130 GeV. 3. The $\kappa(\eta)$ does not strongly 
depend on the centrality. 

A parton and hadron cascade model, PACIAE, is employed in this letter investigating 
the net charge transfer fluctuation, $\kappa$, within $|\eta|<1$ both in the early 
partonic stage (QGM) and in the hadronic final state (HM) in $Au+Au$ collisions at 
$\sqrt{s_{NN}}$=200 GeV. As expected the later results are quite close to the results 
in HIJING, RQMD, and UrQMD. However the former results are smaller than the later one 
by a factor of 3 to 5. Unfortunately, the $\kappa$ in QGM seems to be hard to survive 
the hadronization.      

As the simplified version of PACIAE model has been published in \cite{sa1} and a nice 
bit of detailed description has been given in \cite{sa2}, here we just give a brief 
introduction for PACIAE model. In the PACIAE model a nucleus-nucleus collision 
is decomposed into nucleon-nucleon collisions. The nucleons in a nucleus is 
distributed randomly according to Wood-Saxon distribution. A nucleon-nucleon collision 
is described by the Lowest-Leading-Order (LLO) pQCD parton-parton hard interactions 
with parton distribution function in a nucleon and by the soft interactions considered 
empirically, that is so called "multiple mini-jet production" in HIJING model 
\cite{wang1}. However, in PACIAE model that is performed by PYTHIA model \cite{sjo1} 
with string fragmentation switched-off. Therefore, the consequence of nucleus-nucleus 
collision is a configuration of $q$ ($\bar q$), diquark (anti-diquark), and $g$, 
besides the spectator nucleons and beam remnants \cite{sjo1}. The diquark (anti-
diquark) is forced to split into $qq$ ($\bar q\bar q$) randomly. 

So far we have introduced the partonic initialization of nucleus-nucleus collision in 
PACIAE model, what follows is then parton evolution (scattering). To the end, the 
2$\to$2 LLO pQCD differential cross section \cite{comb} is used. Of course, that must 
be regularized first by introducing the color screen mass. The total cross section of 
parton $i$ bombarding with $j$ could then be calculated via a integral over the 
squared momentum transfer in a subprocess $ij\to kl$ and a summation over partons $k$ 
and $l$. With above differential and total cross sections the parton scattering can be 
simulated by Monte Carlo method. As of now, only 2$\to$2 processes are involved, among 
them there are six elastic and three inelastic processes \cite{comb}.

As for the hadronization we first assume that the partons begin to hadronize when the 
interactions among them have been ceased (freeze-out). They could hadronize by either 
fragmentation model \cite{ff1,and1} or coalescence model \cite{biro1,csiz}. What the 
fragmentation models included here are the Field-Feynman model, i. e. Independent 
Fragmentation (IF) model \cite{ff1} and Lund string fragmentation model \cite{and1}. 
However, the program built in \cite{sjo1} is employed for the implementation of 
fragmentation model. On the contrary, we do write a program for coalescence model 
ourselves.  

The hadron evolution (hadronic rescattering) is modeled as usual two body collisions 
and is copied directly from a hadron and string cascade model LUCIAE \cite{sa3}. There 
is no need to say more about the hadronic rescattering referring to \cite{sa3} if 
necessary.

\vspace{2.5cm}   
\begin{figure}[ht]
\centerline{\hspace{-0.5in}
\epsfig{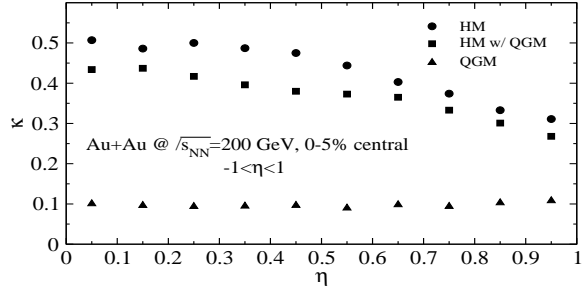}}
\vspace{0.2in}
\caption{Observable of net charge transfer fluctuation, $\kappa$, as a function of 
         $\eta$ within $|\eta|<1$ in 0-5\% most central $Au+Au$ collisions at $\sqrt
         {s_{NN}}$=200 GeV.}
\label{ctrfl1}
\end{figure}

Since we are not aimed to reproduce the experimental date but to study the physics of 
net charge transfer fluctuation, we do not adjust model parameters at all. In the 
calculations the IF model \cite{ff1} is adopted for hadronization and the net charge 
transfer fluctuation is counted in the interval of $|\eta|<$1. The simulated results by 
default PACIAE model are indicated with "HM w/ QGM" (HM with QGP assumption), since 
the hadronic final state is evolved from partonic initial state. If the simulation 
is ended up at the stage of partonic scattering and the net charge transfer 
fluctuation is counted over partons only, the results will be referred to as "QGM". 
In that calculation it is assumed that the gluon does not contribute to the net charge 
but it does contribution to charge multiplicity by 2/3 as assumed in \cite{jeon4,sa4}. 
If the simulation is ended up at the stage of partonic scattering and both the partons 
and beam remnants (hadrons) are counted in net charge transfer fluctuation the 
corresponding results are then symboled as "QGM w/ remnant". It should be mentioned 
here that the spectator nucleons do not affect the net charge transfer fluctuation in 
$|\eta|<$1. A calculation where the string fragmentation in PYTHIA is switched-on and 
followed directly by the hadronic rescattering is referred to as "HM", since in this 
simulation only hadronic transport is taken into account, like in HIJING, RQMD, UrQMD, 
and JPCIAE \cite{sa5}.      

Fig. \ref{ctrfl1} gives the observable of net charge transfer fluctuation, $\kappa$, 
as a function of pseudorapidity, $\eta$, in the simulations of "HM", "HM w/ QGM", and 
"QGM" (solid circles, squares, and triangles, respectively) in 0-5 \% most central 
$Au+Au$ collisions at $\sqrt{s_{NN}}$=200 GeV. One sees in this figure that the 
$\kappa$ of "HM w/ QGM" reproduces nicely the STAR datum and the $\kappa$ of "HM" is a 
bit larger than the STAR datum at $\eta$=1. The trend of $\kappa$ varying with $\eta$, 
both in "HM" and "HM w/ QGM", is similar to the ones in HIJING, RQMD, UrQMD (cf. Fig. 
5 in \cite{jeon2}). On the contrary, the $\kappa$ of "QGM" keeps nearly constant, like 
the charge fluctuation as function of rapidity interval in QGM in thermal model 
\cite{jeon3} and in transport model \cite{sa4}. It is interesting to see that the 
$\kappa$ in "HM" is larger than $\kappa$ in "QGM" by a factor of 3 to 5 from upper  
$\eta$ to the lower $\eta$. The discrepancy between $\kappa$ in "HM" and in "HM w/ QGM
" amounts $\sim$ 20\% in the average, that means that the probability of net charge 
transfer fluctuation in QGM surviving hadronization can be estimated to be $\sim$ 20\% 
either. 

\vspace{2.5cm}
\begin{figure}[ht]
\centerline{\hspace{-0.5in}
\epsfig{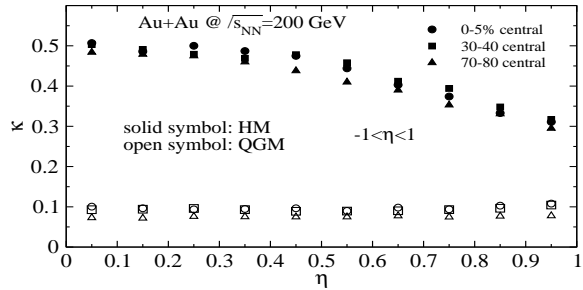}}
\vspace{0.2in}
\caption{$\kappa$ as a function of $\eta$ within $|\eta|<1$ in a range centralities 
         of the $Au+Au$ collisions at $\sqrt{s_{NN}}$=200 GeV.}
\label{ctrfl2}
\end{figure}
  
In Fig. \ref{ctrfl2} the centrality dependence of $\kappa(\eta)$ is given both for 
"HM" and "QGM" (solid and open symbols, respectively) for 0-5, 30-40, and 70-80\% 
central $Au+Au$ collisions at $\sqrt{s_{NN}}$=200 GeV. Both of the $\kappa(\eta)$ in 
"HM" and in "QGM" do not strongly depend on the centrality which is consistent with 
the results in hadronic transport models HIJING, RQMD, and UrQMD (cf. Fig. 5 in 
\cite{jeon2}).

\begin{figure}[ht]
\centerline{\hspace{-0.5in}
\epsfig{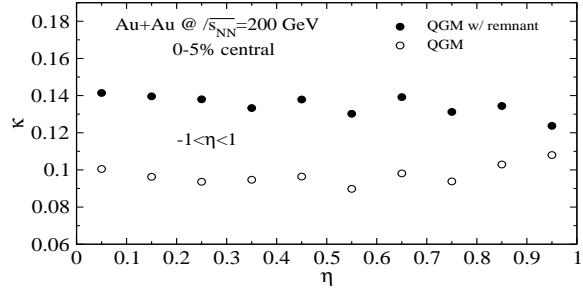}}
\vspace{0.2in}
\caption{$\kappa$ as a function of $\eta$ within $|\eta|<1$ in 0-5\% most central 
         $Au+Au$ collisions at $\sqrt{s_{NN}}$=200 GeV.}
\label{ctrfl3}
\end{figure}

We compare the $\kappa(\eta)$ in "QGM" to the one in "QGM w/ remnant" in Fig. 
\ref{ctrfl3} for 0-5\% most central $Au+Au$ collisions at $\sqrt{s_{NN}}$=200 GeV. One 
sees in this figure that the influence of beam remnants upon the $\kappa(\eta)$ in  
"QGM" amounts $\sim$35\% in average. This influence does not change the status of 
big difference between $\kappa$ in "HM" and in "QGM" shown in Fig. \ref{ctrfl1}. 
    
\vspace{2.5cm}

In summary, a parton and hadron cascade model, PACIAE, is employed investigating the 
net charge transfer fluctuation within $|\eta|$=1 both in partonic initial state and 
in hadronic final state for a range centralities of $Au+Au$ collisions at $\sqrt{s_{NN
}}$=200 GeV. In the hadronic final state ($\kappa$ in "HM") the observable of net 
charge transfer fluctuation, $\kappa$, turns out to be nearly a factor of 3 to 5 
larger than the $\kappa$ in partonic initial state ($\kappa$ in "QGM"). However, the 
$\kappa$ in "QGM" is hard to survive the hadronization, the survival probability 
amounts $\sim$ twenty percent.       

Finally, the financial support from NSFC (10475032) in China are acknowledged.


\begin{thebibliography}{10}
\bibitem{jeon1}
Lijun Shi and S. Jeon, Phys. Rev. C 72, 034904 (2005);
S. Jeon, Lijun Shi, and M. Bleicher, nucl-th/0506025.
\bibitem{jeon2}
S. Jeon, Lijun Shi, and M. Bleicher, nucl-th/0511066.
\bibitem{quig}
C. Quigg and G. H. THomas, Phys. Rev. D 7, 2752 (1973);
A. W. Chao and C. Quigg, Phys. Rev. D 9, 2016 (1974).
\bibitem{jeon3}
S. Jeon and V. Koch, Phys. Rev. Lett. 85, 2076 (2000).
\bibitem{bass1}
S. A. Bass, P. Danielewicz, S. Pratt, Phys. Rev. Lett. 85, 2689 (2000).
\bibitem{asak}
M. Asakawa, U.W. Heinz, B. M$\ddot{u}$ller, Phys. Rev. Lett. 85, 2072 (2000).
\bibitem{wang1}
M. Gyulassy and X.-N. Wang, Comput. Phys. Commun. 83, 307 (1994).
\bibitem{sorg1}
H. Sorge, M.Berenguer, H. St$\ddot{o}$cker, and W. Greiner, Phys. lett. B 289, 6 
(1992).
\bibitem{bass2}
S. A. Bass, et al., Prog. Part. Nucl. Phys. 41, 225 (1998).
\bibitem{star1}
J. Adams, et al., STAR Collaboration., Phys. Rev. C 68, 044905 (2003).
\bibitem{sa1}
Ben-Hao Sa and Aldo Bonasera, Phys. Rev. C 70, 034904 (2004). 
\bibitem{sa2}
Ben-Hao Sa, Dai-Mei Zhou, and Zhi-Guang Tan, J. Phys. G: Nucl. Part. Phys. 32, 243 
(2006).
\bibitem{sjo1}
T. Sj$\ddot{o}$strand, Comput. Phys. Commun. 82, 74 (1994).
\bibitem{comb}
B. L. Combridge, J. Kripfgang, and J. Ranft, Phys. Lett. B 70, 234(1977).
\bibitem{ff1}
R. D. Field and R. P. Feynman, Phys. Rev. D 15, 2590 (1977); Nucl. Phys. 
B 138, 1(1978); R. P. Feynman, R. D. Field, and G. C. Fox, Phys. Rev. D 18, 
3320 (1978).
\bibitem{and1}
B. Andersson, G. Gustafson, G. Ingelman, T. Sj$\ddot{o}$strand, Phys. Rep. 97, 
33 (1983); B. Andersson, G. Gustafson, B. S$\ddot{o}$derberg, Nucl. Phys. 
B 264, 29 (1986).
\bibitem{biro1}
T. S. Biro, P. Levai, and J. Zimanyi, Phys. Lett. B 347, 6 (1995).
\bibitem{csiz}
P. Csizmadia, P. Levai, S. E. Vance, T. S. Biro, M. Gyulassy, and J. Zimanyi, 
J. Phys. G 25, 321 (1999).
\bibitem{sa3}
Sa Ben-Hao and Tai An, Comput. Phys. Commun. 90, 121 (1995);
Tai An and Sa Ben-Hao, Comput. Phys. Commun. 116, 353 (1999).
\bibitem{jeon4}
S. Jeon and V. koch, Phys. Rev. Lett. 83 5435 (2000).
\bibitem{sa4}
Ben-Hao Sa, Xu Cai, An Tai, and Dai-Mei Zhou, Commun. Theor. Phys. (Beijing, China)
42 403 (2004).
\bibitem{sa5}
Ben-Hao Sa, An Tai, Hui Wang, and Feng-He Liu, Phys. Rev. C 59, 2728 (1999).
\end{thebibliography}
\end{document}